\documentclass[twocolumn,article,amsmath,amssymb,aps,prb,floatfix,showkeys,superscriptaddress]{revtex4-1}
\usepackage{lipsum}
\usepackage{capt-of}
\usepackage{bm}
\usepackage{amssymb}
\usepackage{amsmath}
\usepackage{graphicx}
\usepackage{epsfig}
\usepackage{array}
\usepackage{float}
\usepackage{color}
\usepackage[usenames,dvipsnames]{xcolor}
\usepackage{epstopdf}
\usepackage{natbib}
\usepackage{lineno}
\usepackage{slashed}
\usepackage{color}
\usepackage{hyperref}
\usepackage{slashed}
\usepackage{soul}

\newcommand{\be}{\begin{equation}}
\newcommand{\ee}{\end{equation}}
\newcommand{\bea}{\begin{eqnarray}}
\newcommand{\eea}{\end{eqnarray}}
\newcommand{\beas}{\begin{eqnarray*}}
\newcommand{\eeas}{\end{eqnarray*}}

\DeclareFontFamily{OMS}{oasy}{\skewchar\font48 }
\DeclareFontShape{OMS}{oasy}{m}{n}{%
         <-5.5> oasy5     <5.5-6.5> oasy6
      <6.5-7.5> oasy7     <7.5-8.5> oasy8
      <8.5-9.5> oasy9     <9.5->  oasy10
      }{}
\DeclareFontShape{OMS}{oasy}{b}{n}{%
       <-6> oabsy5
      <6-8> oabsy7
      <8->  oabsy10
      }{}
\DeclareSymbolFont{oasy}{OMS}{oasy}{m}{n}
\SetSymbolFont{oasy}{bold}{OMS}{oasy}{b}{n}

\DeclareMathSymbol{\smallleftarrow}     {\mathrel}{oasy}{"20}
\DeclareMathSymbol{\smallrightarrow}    {\mathrel}{oasy}{"21}
\DeclareMathSymbol{\smallleftrightarrow}{\mathrel}{oasy}{"24}

\newcommand{\tens}[1]{\overset{\scriptscriptstyle\smallleftrightarrow}{#1}}

\begin{document}

\title{Critical behavior for point  monopole and dipole electric impurities in uniformily and uniaxially strained graphene }

\author{Julio C\'esar P\'erez-Pedraza}
\affiliation{                    
 Instituto de F\'{\i}sica y Matem\'aticas, Universidad Michoacana de San Nicol\'as de Hidalgo. Edificio C-3, Ciudad Universitaria. Francisco J. M\'ujica s/n. Col. Fel\'{\i}citas del R\'{\i}o. CP 58040, Morelia, Michoac\'an, Mexico.}
 \author{Erik D\'{\i}az-Bautista}
 \affiliation{Departamento de Formaci\'on B\'asica Disciplinaria, Unidad Profesional Interdisciplinaria de Ingenier\'ia Campus Hidalgo del Instituto Polit\'ecnico Nacional, Pachuca: Ciudad del Conocimiento y la Cultura, Carretera Pachuca-Actopan km 1+500, San Agust\'in Tlaxiaca, 42162 Hidalgo, M\'exico}
 \author{Alfredo Raya}
 \affiliation{                    
 Instituto de F\'{\i}sica y Matem\'aticas, Universidad Michoacana de San Nicol\'as de Hidalgo. Edificio C-3, Ciudad Universitaria. Francisco J. M\'ujica s/n. Col. Fel\'{\i}citas del R\'{\i}o. CP 58040, Morelia, Michoac\'an, Mexico.}
 \affiliation{                    
 Centro de Ciencias Exactas, Universidad del B\'{\i}o-B\'{\i}o, Avda. Andr\'es Bello 720, Casilla 447, 3800708, Chill\'an, Chile}
 \author{David Valenzuela}
\affiliation{Instituto de F\'{\i}sica, Pontificia Universidad Cat\'olica de Chile, Casilla 306, Santiago 22, Chile}

\begin{abstract}
    We revisit the problem of bound states in graphene under the influence of point electric monopole and dipole impurity potentials extended to the case in which the membrane of this material is uniformly and uniaxially strained,
    which leads to a redefinition of the charge and dipole moment, respectively. By considering an anisotropic Fermi velocity, we analytically solve the resulting Dirac equation for each potential. We observe that the effect of the anisotropy is to promote or inhibit the critical behavior known to occur for each kind of impurity, depending on the direction along which strain is applied: both the atomic collapse, for the monopole impurity, and the emergence of cascades of infinitely many bound states with a universal Efimov-like scaling, for the dipole impurity, are phenomena that occur under less or more restrictive conditions due to strain. 
\end{abstract}

\maketitle

\section{Introduction}

Ever since the isolation of graphene samples in 2004 \cite{novoselov2004electric}, this material has played a leading role in exploring analogs and similarities of phenomena occurring in different branches of physics. In particular, the low-energy quasi-particle excitations in this material, which behave as ultra-relativistic Dirac fermions in two space dimensions, have permitted to establish long standing predictions in quantum electrodynamics (QED) but in a condensed matter/solid state realm. Klein tunneling~\cite{katsnelson2006chiral, stander2009evidence, young2009quantum}, Zitterbewegung~\cite{katsnelson2006zitterbewegung, auslender2007generalized} and atomic collapse~\cite{shytov2007vacuum, shytov2007atomic, pereira2007coulomb, novikov2007elastic} are examples of phenomena that have been experimentally observed and theoretically predicted to occur in graphene due to the pseudo-relativistic character of its charge carriers near the Dirac points. In QED,  the phenomenon of atomic collapse refers to the prediction that an electron would dive into the nucleus of a super-heavy element emitting a positron when the charge of the nucleus exceeds certain critical threshold, $Z_c$~\cite{pomeranchuk1945energy, greiner1985quantum}. The physical explanation supporting this unobserved prediction is that intense Coulomb field causes the electron wave function component to fall into the center of the nucleus (with a finite lifetime~\cite{zeldovich1972electronic}), and, at the same time, the positron component escapes to infinity. Such supercritical quantum states depict the semiclassical idea of an electron spiralling inward toward the nucleus as the corresponding positron spirals outward away from it, propagating to infinity. For a punctual nucleus, the threshold is calculated to be $Z_c=137$, whereas for extended nucleus, $Z_c\simeq 170$~\cite{pomeranchuk1945energy, greiner1985quantum, zeldovich1972electronic}.  Up to date, this phenomenon has not been observed in nuclear or heavy-ion experiments nor in any other physical context.

In graphene, it has been observed by several groups that the low-energy Dirac Hamiltonian becomes non self-adjoint when the coupling of the long-range static Coulomb interaction  for $\beta  = Z\alpha/\kappa> \beta_c \simeq 1/2$ \cite{shytov2007vacuum, shytov2007atomic, pereira2007coulomb, novikov2007elastic, terekhov2008screening, valenzuela2016atomic}. Here, $\alpha$ is the fine structure constant, $\kappa$ the dielectric constant and $Z$ the charge of an artificial nucleus. In this material, the critical charge $Z_c$ of {\it artificial nuclei} (Coulomb impurity) is strongly reduced \cite{shytov2007vacuum, shytov2007atomic, pereira2007coulomb, valenzuela2016atomic, fogler2007screening}. This theoretical prediction has been experimentally confirmed~\cite{wang2013observing} by using the tip of a scanning tunnelling microscope (STM) to create clusters of charged calcium dimmers, creating a supercritically charged Coulomb potential from subcritical charges (as in heavy-ion collisions).
In these experiments,
a systematic electron-hole asymmetry owing to the positive charge of the dimmers was observed, along with the appearance of a resonance (a quasi-bound state) increasing and shifting toward lower energies as the number of dimmers increased from one to five.  For five dimmers, the resonance shifted below the Dirac point, representing the atomic collapse of the electron in graphene.

As a natural generalization, the problem of graphene under the influence of an electric dipole impurity has recently been addressed by several groups~\cite{klopfer2014scattering, de2014electric, gorbar2015supercritical, van2018electrical, cuenin2014dipoles}. For a finite-size electric dipole impurity, it has been observed that when each of the charges  of the  dipole exceeds a critical value, the wavefunction of the highest energy occupied bound state changes its localization from the negatively charged impurity to the positive one as the distance between the impurities changes. This phenomenon has been dubbed as the  migration of the electron wave function and comprises a generalization of the atomic collapse phenemonon by a single charged imputity  to the scenario where both electrons and holes are spontaneously created from the vacuum in bound states with the two charges of the dipole partially screening them.
Nevertheless, in the point electric dipole limit, which occurs in the limit when the distance between the charges of the dipole vanishes with fixed electric dipole moment, a reminiscent of the critical behavior is observed: there exists at least one cascade of infinitely many bound states for an arbitrary value of the dipole strength. Additional cascades develop when the dipole strength achieves critical values according to a universal Efimov-like scaling \cite{efimov1970energy, braaten2006universality, gogolin2008analytical}. There is no collapse of the electron wave function to the continuum because even though the electron and positron energy levels tend to come close to each other as the dipole moment increases, eventually they separate apart, thus preventing collapse.

In this article we revisit these critical problems in uniformly and uniaxially strained graphene. We consider the case in which the sample is first strained and then the impurities are later located in the plane. Mechanical deformations of the graphene membranes are known to modify the electronic properties of the material~\cite{pereira2009tight, choi2010effects, farjam2009comment, ribeiro2009strained, cocco2010gap, guinea2010energy, guinea2010generating}. As a matter of fact, straintronics~\cite{sahalianov2019straintronics} has emerged as the field of research studying the modifications of graphene properties due to strain. For the simplest case we address in this article, it is well known that a uniform and uniaxial strain still induces an anisotropy in the Fermi velocity that, nevertheless, does not generate pseudomagnetic fields~\cite{guinea2010energy}. 
Thus, by solving the corresponding anisotropic Dirac equation under the influence of monopole and dipole electric impurity potentials, we explore the role of the anisotropy on the critical thresholds for these problems. For this purpose, we have organized the remaining of this article as follows: In the next Section we solve the anisotropic Dirac equation for the case of a monopole electric impurity immersed in a uniform magnetic field~\cite{gamayun2011magnetic} and explore the behavior of the bound state energies. We regularize the Coulomb potential at the origin replacing it with a potential well extended up to a finite ``elliptic radius'' (this point will become clear in the discussion below) $r_0$ and of strength $V_0$. We explore the effect of the anisotropy parameter in the collapse of bound states at different angular momenta for several regimes. In Section 3 we address the problem of a point electric dipole in uniaxially strained graphene. We observe the impact of the strain parameter in the emergence of towers of infinitely many bound states and the changes that it induces in the Efimov-like scaling of these cascades of bound states of the pristine case. We discuss our findings and present our conclusions in Section 4. An appendix with useful formulae is also included.

\section{Monopole impurity and a magnetic field in strained graphene}

We start from the 2D Dirac Hamiltonian describing an electric monopole  impurity in graphene immersed in a perpendicular magnetic field $\textbf{B}$ uniformly distributed in space. Considering the tensor character of the Fermi velocity induced by strain,  $\tens{v}_{\rm F}= v_{\rm F}\,{\rm diag}(a,b)$~\cite{betancur2015landau, concha2019barut, diaz2019coherent, diaz2020wigner}, such an equation is written as
\begin{equation}
H_{\rm D} = \hbar\ \tensor{v}_{\rm F}\cdot \boldsymbol{\sigma}\cdot \boldsymbol{\pi}+\xi\Delta\sigma_3+V(r'),
\end{equation}
\noindent where $V(r')$ corresponds to impurity potential under strain, $a$, $b$ are the strain parameters along each spatial direction, $\pi_i = -i \partial_i + (e/\hbar c) A_i$ with $i=x,y$, is the $i$-th component of the canonical momentum, $\mathbf{A} =\frac{B_0}{2}(-y,x)$ is the symmetric gauge vector potential describing the external magnetic field $\mathbf{B}$, $\sigma_i$ are the Pauli matrices, and $\Delta$ is the quasiparticle mass gap. The two component spinor $\Psi_{\xi s}$ carries the valley ($\xi= \pm$) and spin ($s=\pm$) indices. Also, the standard convention $\Psi_{+s} = (\psi_A,\psi_B)_{K^T_{+s}}$ whereas $\Psi_{-s} = (\psi_B,\psi_A)_{K^T_{-s}}$ is used. Here, $A$, $B$ refer to the two sublattices of a hexagonal graphene crystal. Because the interaction $V(r')$ does not depend on spin, this quantum number is omitted throughout. Then, by defining the parameters
\begin{equation}\label{parameters}
    \lambda = \sqrt{ab},\qquad \mbox{and} \qquad \zeta = \frac{a}{b} \ ,
\end{equation} 
the equation governing the system can be written as
\begin{equation}
\left(
\begin{array}{cc}
0 & -iS^- \\
i S^+ & 0
\end{array}
\right)
\Psi(\mathbf{r}) =\left( \begin{array}{cc}
\epsilon^{-}  & 0 \\
0 & \epsilon^{+}
\end{array}\right)\Psi(\mathbf{r}),\label{eq:start}
\end{equation}\\
 where we have defined the operators
 \begin{equation}
     S^{\pm} = \mp i \frac{l_{\rm B}}{\sqrt{2}} \left( \zeta^{1/2}\pi_x\pm i\zeta^{-1/2}\pi_y \right),\ \ [ S^{-},  S^{+}] = 1\ ,\label{operators}
 \end{equation}
 and the quantities\\
 \begin{equation}
\epsilon^{\pm} = \frac{l_{\rm B}[E-V(r')\pm \xi\Delta]}{\sqrt{2}\lambda \hbar v_{\rm F}}\ .
\end{equation}\\
Here, $l_{\rm B}=\sqrt{\hbar c/|eB_0|}$ is the magnetic length. 
Defining the following azimuthal-symmetric relations
\begin{equation}\label{transform}
    x=\zeta ^{1/2}\ r\ \cos\phi, \qquad y = \zeta ^{-1/2}\ r\ \sin\phi,
\end{equation}
which describe the ellipse\\
\begin{equation}
    \frac{x^2}{\zeta r^2}+\frac{y^2}{\zeta^{-1} r^2}= 1 \ ,\label{radial}
\end{equation}
we have that
\begin{equation}
\zeta ^{1/2} \pi_x\pm i\zeta ^{1/2}\pi_y ={\rm e}^{\pm i\phi} \left( -i\ \partial_r \pm \frac{\partial_{\phi}}{r}  \pm \frac{r}{2l_{\rm B}^2} \right),
\end{equation}
hence obtaining the following form of the operators in~(\ref{operators}),
\begin{equation}
    S^{\pm}= \mp \frac{1}{\sqrt{2}}{\rm e}^{\pm i\phi} \left( \ \partial_{\rho} \pm \frac{i\ \partial_{\phi}}{\rho}  \mp \frac{\rho}{2} \right),
\end{equation}
with the substitution $\rho = r/l_{\rm B}$ in the last expression. For further details about how the dimensionless quantities $a$, $b$ (Eq.~(\ref{parameters})) depend on the lattice parameters and the tensile strain $\varepsilon$, that measures compress and tensile deformations on the layer, the reader is referred to~\cite{betancur2015landau, diaz2020wigner}. In addition, the transformation showed in Eq.~(\ref{transform}) allows us to solve the problem in a new coordinate system in which the $\zeta$-dependency remains implicitly. More precisely, $r^{2}=\zeta^{-1}x^{2}+\zeta\ y^{2}$ and $\tan(\phi)=\zeta\ y/x$. On the other hand,  considering strain in the Coulomb potential ($q/r'$), with $r'^2=\zeta r^2 \cos^2 \phi + \zeta^{-1}r^2 \sin^2\phi$ from Eq.~(\ref{radial}), we found that
\begin{align}
\dfrac{r}{r'} &= \zeta^{-1/2}\left[ 1 + (\zeta^{-2}-1) \sin^2 \phi \right]^{-1/2} \nonumber \\ &\approx \zeta^{-1/2}\left[1- \dfrac{1}{2} (\zeta^{-2}-1) \sin^2 \phi\right],
\end{align} 
thus, obtaining the zero-order approximation of the strained Coulomb potential
\begin{equation}
V_{\rm Coul}(r')= \dfrac{q}{r'} \approx \dfrac{q}{r}\zeta^{-1/2} = \dfrac{q'}{r}=V_{\rm Coul}(r) ,
\end{equation}
with the charge redefinition $q' = q \zeta^{-1/2}$. Unfortunately, it is well known that the problem of the Dirac equation in a magnetic field plus a Coulomb potential cannot be solved exactly~\cite{gamayun2011magnetic}. Thus, considering a finite-size monopole impurity and neglecting the Coulombic tail (to avoid the difficulties of the analytical examination of this problem), we regularize the Coulomb potential at the origin by considering a potential well $V(r')=-V_0 \theta (r'_0-r')$, including the effects of strain in the radial variable $r$. The argument of the Heaviside function
\begin{align}
r'_0-r'&= \left[ \zeta r_0^2 \cos^2\phi + \zeta^{-1} r_0^2 \sin^2\phi\right]^{1/2} \nonumber \\ &\quad- \left[ \zeta r^2 \cos^2\phi + \zeta^{-1} r^2 \sin^2\phi\right]^{1/2} \nonumber \\\nonumber \\ &\approx \zeta^{1/2} (r_0-r)\left[1+ \dfrac{(\zeta^{-2}-1)}{2} \sin^2\phi\right].
\end{align}
Using Eq.~(\ref{eq:heaviside}) of the Appendix, and noting that $\zeta^{1/2}\left(1+ (\zeta^{-2}-1) \sin^2\phi/2\right) >0$ ($a,b>0$), we obtain 
\begin{equation}
    V_{\rm well}(r') = -V_0\theta(r'_0-r') \approx -V_0 \theta(r_0-r)=V_{\rm well}(r), \label{eq:well}
\end{equation}
so that, the potential well remains invariant under strain. Then,  from Eq.~(\ref{eq:start}) we get 
\begin{equation}
S^-S^+ \psi_A = \epsilon^+ \epsilon^- \psi_A,\qquad S^+S^- \psi_B =\epsilon^+ \epsilon^- \psi_B,\label{decoupled}
\end{equation}
where
\begin{equation}
\epsilon^+ \epsilon^- = p'^2=\left\{
\begin{array}{cc}
p_v^2= \displaystyle{\frac{l_{\rm B}^2[(E+V_0)^2-\Delta^2]}{2(\lambda \hbar v_{\rm F})^2}}, & \rho<\rho_0 \\ 
p^2=\displaystyle{\frac{l_{\rm B}^2[E^2-\Delta^2]}{2(\lambda \hbar v_{\rm F})^2}}, & \rho>\rho_0 
\end{array}\right. .\label{def-ps}
\end{equation}
The spinor $\Psi(\mathbf{r})$ can be written in terms of the eigenfunctions of the conserved angular momentum $J_z = L_z + \sigma_z/2 = -i \partial_{\phi} + \sigma_z/2$ as follows:
\begin{equation}
\Psi(\mathbf{\rho}) = \frac{1}{\rho} \left(
\begin{matrix}
{\rm e}^{i\left( j-1/2\right) \phi}\ f(\rho) \\ \\
i\ {\rm e}^{i\left( j+1/2\right) \phi}\ g(\rho)
\end{matrix} \right) ,
\end{equation}
with $j=\pm 1/2$, $\pm 3/2$,... Substituting this spinor in Eq.~(\ref{decoupled}), through the replacement $x=\rho^2/2$, we have for the upper component
\begin{equation}
f''+\left[\frac{p'^2-j/2-1/4}{x}+\frac{1/4-(j/2-1/4)^2}{x^2}-\frac{1}{4} \right] f =0 ,
\end{equation}
which correspond to a Whittaker differential equation, whose general solution is expressed in terms of confluent hypergeometric functions $M(a,c;x)$ and $U(a,c;x)$~\cite{bateman1953higher} as
\begin{align}
f(x)= & {\rm e}^{-\frac{x}{2}} x^{\frac{j}{2}+\frac{1}{4}} \left[C_1 M\left(j+\frac{1}{2}-p'^2,j+\frac{1}{2};x\right)+ \right. \nonumber \\ \nonumber \\ & \left. C_2 U\left(j+\frac{1}{2}-p'^2,j+\frac{1}{2};x\right)\right].
\end{align}
From this solution, we can obtain the general expression for $g(\rho)$ by acting with $S^+$ upon the upper component of the spinor $\psi_A$. Moreover, from the asymptotic behavior of $M(a,c;x)$ and $U(a,c;x)$, 
the regular solutions at the origin $\rho\to 0$ (int) and $\rho\to\infty$ (ext) for each component are, respectively,
\begin{align}
f_{\rm int}(\rho)&= {\rm e}^{-\frac{\rho^2}{4}} \rho^{c}\frac{C_1}{\Gamma(c)}  M\left(c-p_v^2,c;\frac{\rho^2}{2}\right),\\
f_{\rm ext}(\rho)&= {\rm e}^{-\frac{\rho^2}{4}} \rho^{c}\ C_2 U\left(c-p^2,c;\frac{\rho^2}{2}\right),\\
g_{\rm int}(\rho)&= \frac{l_{\rm B}(E+V_0-\xi\Delta)}{\sqrt{2}\lambda \hbar v_{\rm F}}{\rm e}^{-\frac{\rho^2}{4}} \rho^{c+1}\frac{C_1}{\Gamma(c+1)}\nonumber\\
&\times M\left(c-p_v^2,c+1;\frac{\rho^2}{2} \right),\\
g_{\rm ext}(\rho)&= \frac{\sqrt{2}\lambda \hbar v_{\rm F} C_2}{l_{\rm B}(E+\xi\Delta)}{\rm e}^{-\frac{\rho^2}{4}} \rho^{c+1} \nonumber\\ &\times U\left(c-p^2,c+1;\frac{\rho^2}{2}\right),
\end{align}
where $C_1$ and $C_2$ are arbitrary constants, $\Gamma(n)$ denotes the Gamma function~\cite{bateman1953higher} and $c=j+1/2$. These expressions are valid for $j=\pm1/2$, $\pm3/2$, $\pm5/2$, $\ldots$ Continuity of the spinors at the boundary imply that
\begin{equation}
\left. \frac{f_{\rm int}(\rho)}{f_{\rm ext}(\rho)}\right\vert_{\rho=\rho_0} = \left. \frac{g_{\rm int}(\rho)}{g_{\rm ext}(\rm \rho)}\right\vert_{\rho=\rho_0} ,
\end{equation}
condition that prescribes  a transcendental relation for the energy eigenvalues of the system with total angular momentum $j$. Explicitly:
\begin{align}
\frac{2(\lambda\hbar v_{\rm F})^2 (j+\frac{1}{2})M\left(j+\frac{1}{2}-p_v^2,j+\frac{1}{2};\rho_0^2/2\right)}{l_{\rm B}^2(E+V_0-\xi\Delta)M\left(j+\frac{1}{2}-p_v^2,j+\frac{3}{2};\rho_0^2/2\right)  }& \nonumber\\
&\hspace{-5.5cm}= (E+\xi\Delta) \frac{U\left(j+\frac{1}{2}-p^2,j+\frac{1}{2};\rho_0^2/2\right)} {U\left(j+\frac12-p^2,j+\frac{3}{2};\rho_0^2/2\right)}. \label{eqg}
\end{align}
 This equation reduces to the corresponding relation for the unstrained case  considered in~\cite{gamayun2011magnetic} when $\lambda=1$. In fact, a straightforward analysis follows merely replacing $v_{\rm F}\to \lambda v_{\rm F}$. Notice that because
the functions $M(a,c;x)$ and $U(a,c;x)$ are not well defined for $c=0$, $-1$, $-2$, $\ldots$, which in our case correspond to  negative values of $j$, by using the identities~(\ref{eq:MU}) and (\ref{eq:MUG}) of the Appendix, we are lead to the alternative expression valid  for $j\leq-1/2$: 
\begin{align}
\left(\frac{E}{\Delta}+\frac{V_0}{\Delta}+\xi\right)\left(\frac{1}{\frac{1}{2}-j}\right) \frac{M\left(1-p_v^2,\frac{3}{2}-j;\rho_0^2/2\right)}{M\left(-p_v^2,\frac{1}{2}-j;\rho_0^2/2\right)} &=\nonumber\\
&\hspace{-6cm}-\left(\frac{E}{\Delta}+\xi\right) \frac{U\left(1-p^2,\frac{3}{2}-j;\rho_0^2/2\right)} {U\left(-p^2,\frac{1}{2}-j;\rho_0^2/2\right)}, \label{eqn}
\end{align}
which we analyze below in some relevant regimes.

\subsection{Collapse without external magnetic field}

\begin{figure}[th!]
   \includegraphics[width=\columnwidth]{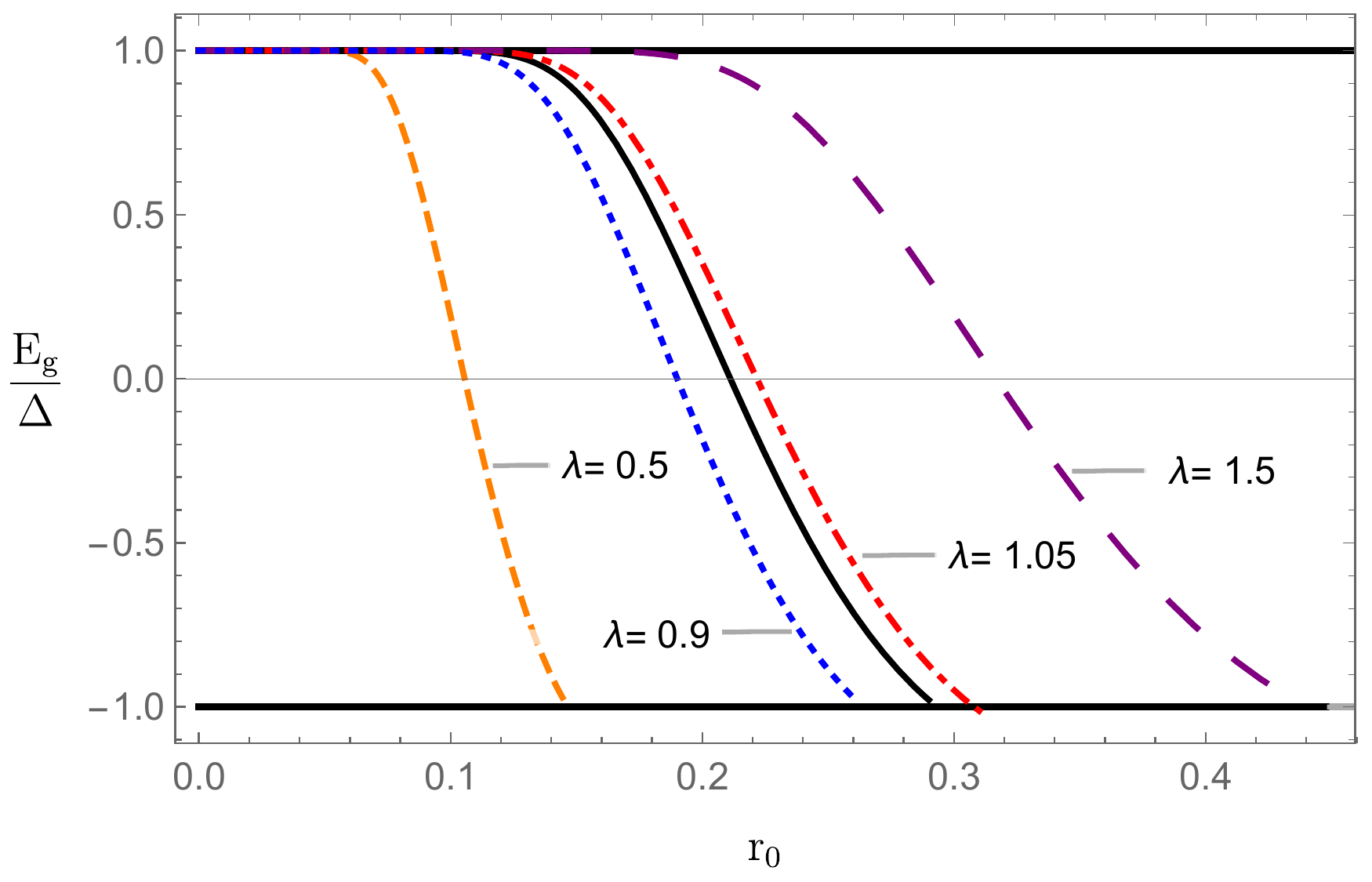}
   \caption{Ground state energy $E_g$ as a function of $r_0$ for different values of $\lambda$ for the state with $j=-1/2$. The unstrained case $\lambda=1$ is indicated by the solid (black) curve. Atomic collapse is promoted for $\lambda<1$ and inhibited for $\lambda>1$.}
    \label{Fig1}
\end{figure}

Let us first analyze the case of the potential well in graphene without the external magnetic field ($B=0 \Rightarrow l_{\rm B}\rightarrow \infty \Rightarrow p_v^2\rightarrow \infty$). By using the identities in Eq.~(\ref{eq:MUB0}) of the Appendix and taking $z= p_v^2 \left(\rho_0^2/2\right) \Rightarrow \rho_0^2/2 = z/p_v^2$, Eq. (\ref{eqg}) becomes
\begin{align}
\frac{ \sqrt{(E+V_0)^2-\Delta^2} J_{j-1/2}\left(\beta r_0\right)}{(E+V_0-\xi\Delta)\  J_{j+1/2}\left(\beta r_0 \right)} &=\nonumber\\ 
   &\hspace{-3cm}
   \frac{\sqrt{E^2-\Delta^2} H^{(1)}_{j-1/2}\left(\beta ' r_0\right)} {(E-\xi\Delta)H^{(1)}_{j+1/2}\left(\beta ' r_0\right)} ,\label{enB0}
\end{align}
with 
\begin{equation}
\beta = \frac{\sqrt{(E+V_0)^2-\Delta^2}}{\hbar \lambda v_{\rm F}}, \qquad \beta ' = \frac{\sqrt{E^2-\Delta^2}}{\hbar \lambda v_{\rm F}}. \label{betas}
\end{equation}
Without loss of generality, let us focus on valley $K_-$ ($\xi=-1$)~\footnote{The solution at the $K_+$ valley is obtained by changing $\Delta \rightarrow -\Delta$ and exchanging the spinor components $\psi_A \leftrightarrow \psi_B$.}. The energy spectrum consists of a continuum part for $|E|> \Delta$, and a discrete one for $|E|< \Delta$, corresponding to bound states. The first bound state, $E_g\lesssim \Delta$, is associated to the smallest centrifugal barrier $j=-1/2$, and appears at an arbitrarily small potential strength $V_0$. Under these considerations, the quantities $\beta$ and $\beta '$ in Eq.~(\ref{betas}) become purely imaginary. 
Using the relations of the Bessel functions in Eqs.~(\ref{Bessel}) and (\ref{Besselz0})
into Eq.(\ref{enB0}),
we have 
\begin{equation}
\frac{E_g}{\Delta} = \left[1- 2 \left(\frac{\hbar \lambda v_{\rm F}}{\Delta r_0} \right)^2\ \text{exp}\left( - \frac{2(\hbar \lambda v_{\rm F})^2}{V_0 \Delta r_0^2} - 2\gamma_E \right)\right],
\end{equation}
where $\gamma_E$ is the Euler-Mascheroni constant ($\gamma_E\approx 0.5772)$. The dependence of the ground state energy with respect to size of the monopole impurity $r_0$ at fixed $V_0$ is shown in Fig.~\ref{Fig1}. In the unstrained case, as the size of the impurity increases, the ground-state energy lowers till $r_0$ reaches a critical size where $E_g$ reaches $E_g=-\Delta$, causing collapse. When we turn on the strain such that $\lambda<1$, collapse occurs for impurities of smaller size. The quoted value $\lambda=$0.9 is consistent with the reported values of strain for which the graphene membrane can be contracted before folding~\cite{contreras2019propagation}. According to~\cite{betancur2015landau,diaz2020wigner}, this value for $\lambda$ would obtain by tensile deformations around $\varepsilon=10\%$ either along the zigzag direction or the armchair direction. For $\lambda>1$, impurities need to be larger in order for strain to drive the collapse. The value $\lambda=$1.05 corresponds to compression deformation of 10\% either along the zigzag direction or the armchair direction. It is worth to mention that the maximum reported value of strain before the crystal breaks corresponds to $\lambda=1.25$\ \cite{naumis2017electronic}.
%

\begin{figure}[th!]
   \includegraphics[width=\columnwidth]{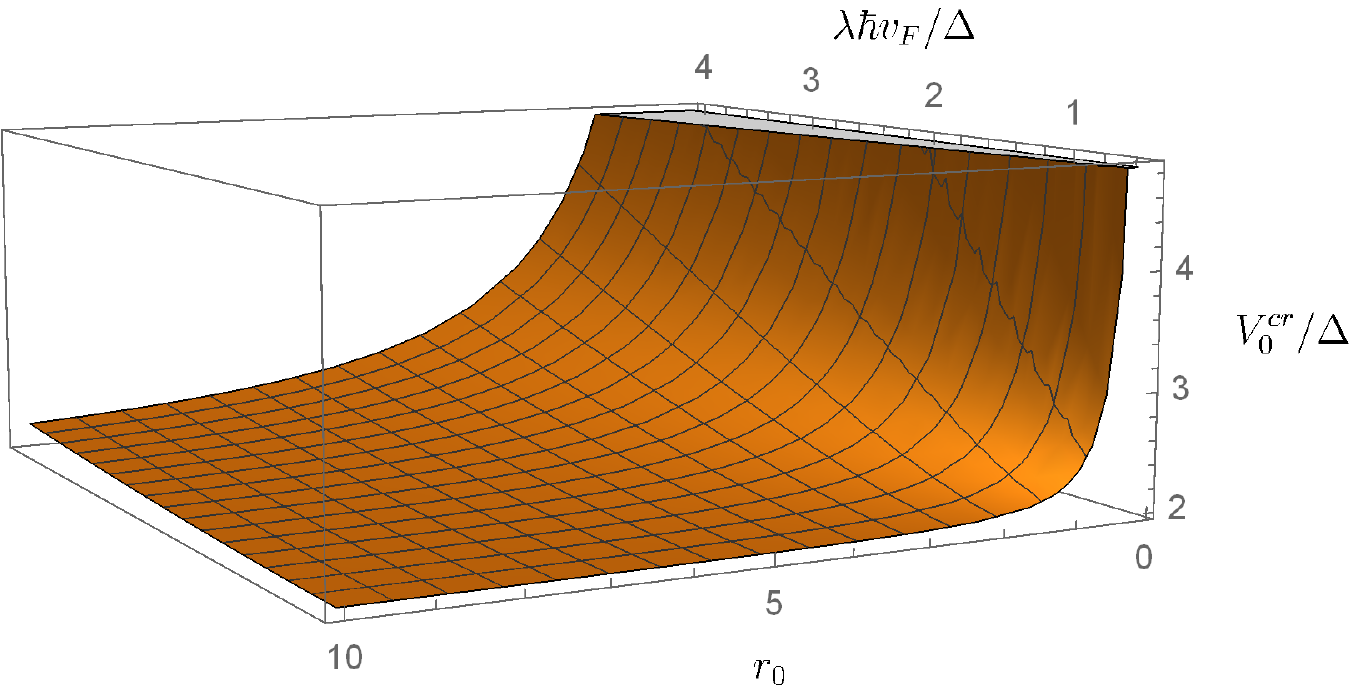}
   \caption{3D plot showing the dependence of $V_{0}^{\rm cr}$ with respect to the variables $r_0$ and $\lambda$.}
    \label{fig:red}
\end{figure}

On the other hand, as $V_0$ grows, the ground state gets closer to  the threshold $E=-\Delta$. In Fig.~\ref{fig:red}, we show the surface of the parameters $\lambda v_{\rm F}$, $r_0$ and $V_0/\Delta$ for which $E_g=-\Delta$. Crossing occurs at the critical value of $V_0^{\rm cr}$,
\begin{equation}
    V_{0}^{\rm cr} = \Delta \left[ 1 + \sqrt{1+\left(\frac{\hbar \lambda v_{\rm F}}{\Delta r_0}\right)^2\ j_{0,1}^2} \right],\label{V0crB0}
\end{equation}
where $j_{0,1} \approx 2.41$ is the first zero of the function $J_0(x)$~\cite{abramowitz1948handbook}. From Eq.~(\ref{V0crB0}), it can be seen  that values of $\lambda >1$ lead to a higher value of $V_{0}^{\rm cr}$, whereas for  $\lambda <1$, this threshold is smaller. We can think of this observation, as the strain deforms the lattice, the Dirac cones have now an elliptical cross section. Thus, the electrons orbiting near the larger semi-axis last more in their spiral-falling trajectory than those that are near the lower semi-axis.  

\subsection{Collapse including magnetic field}

\begin{figure}[th!]
   \includegraphics[width=\columnwidth]{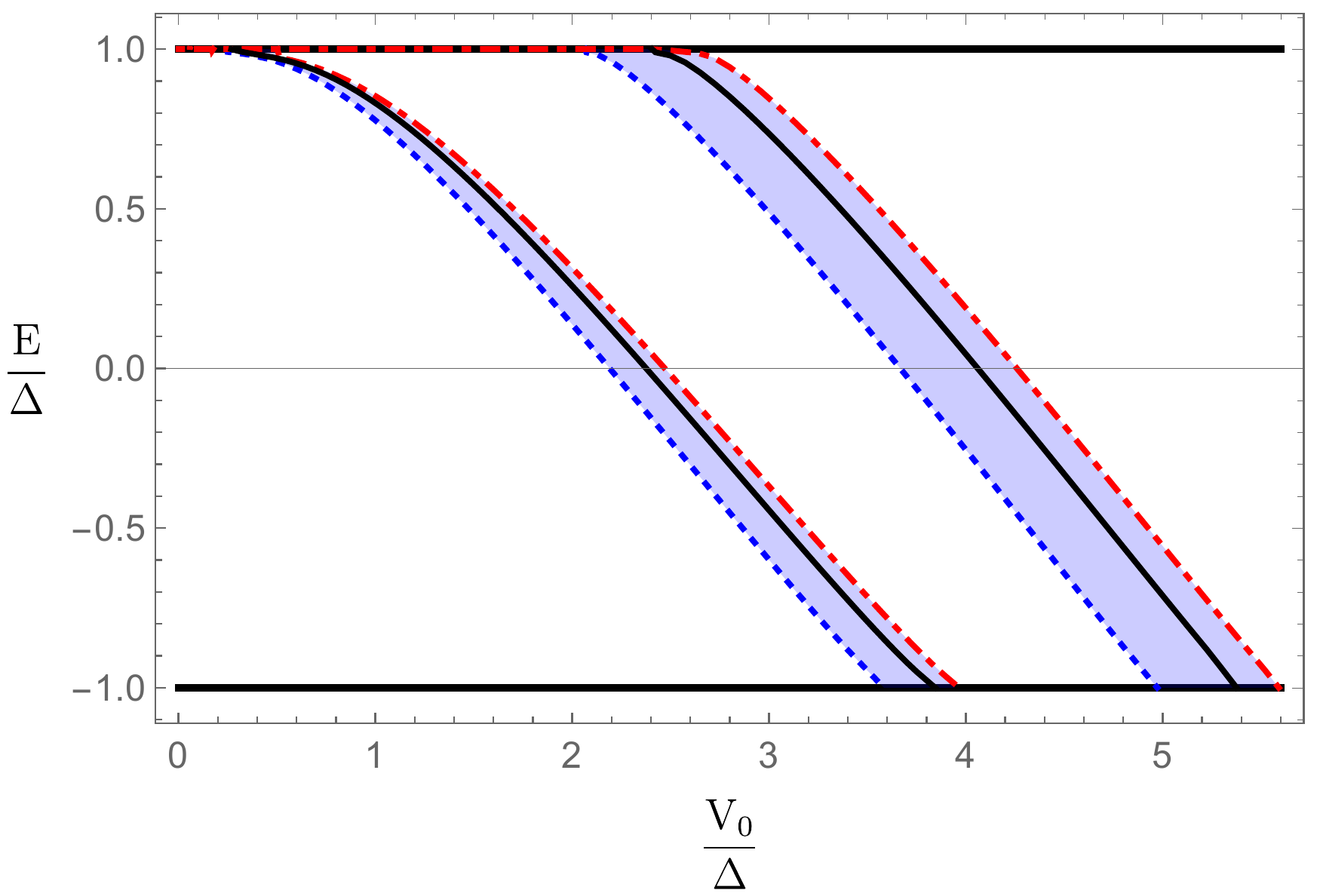}
   \caption{Energy eigenvalue $E$ as a function of $V_0$ (both normalized by the mass gap $\Delta$) for $j=-1/2$ (left strip) and $j=-3/2$ (right strip). The width of these strips corresponds to the extreme values $\lambda= 0.9$ and $\lambda = 1.05$.} 
    \label{fig:energies_v_0_l}
\end{figure}

Let us now switch-on the effect of the external magnetic field. First, notice that neglecting the potential well potential, Eq. (\ref{eqg}) in the $K$ point and with $j \leq -1/2$ render the energy eigenvalues in the well-known form of LLs corrected by strain through the renormalization of the Fermi velocity $v_{\rm F}\to \lambda v_{\rm F}$~\cite{betancur2015landau, concha2019barut, diaz2019coherent, diaz2020wigner}, namely,
\begin{equation}
    E_n = \pm \sqrt{\Delta^2 +2n \left( \frac{\hbar \lambda v_{\rm F}}{l_{\rm B}}\right)^2}\ , \quad n=1,2,...,
\end{equation}
and $j+1/2 \leq n$.
It is important to mention that these states are degenerate because the energy only depends on the principal number $n$ and not on the angular momentum value $j$. This degeneracy is lifted when a nonzero potential $V_0$ is considered. In Fig~\ref{fig:energies_v_0_l}, we plot the energy eigenvalues for $j=-1/2$ and $j=-3/2$ for the extreme values of strain $\lambda=$0.9 and $\lambda=$1.05, both for the valley $\xi=-$. 
There are some important facts to notice: first, as we said before, the states with different angular momentum are no longer degenerate, and more and more of them decay from $E= \Delta$ to $E=-\Delta$ as $V_0$ grows. This represents the fact that these states are diving into the (artificial) nucleus, taking place the atomic collapse. In addition, comparing with the case of unstrained graphene~\cite{gamayun2011magnetic}, an expansive strain inhibits the atomic collapse from appearing, whereas an contracting one promotes collapse. 

The first crossing to the continuum of energy is given by the state with $j=-1/2$. The critical value of the potential $V_{0}^{\rm cr}$ in which this occur (when $E_g=-\Delta$) can be obtained from Eq.~(\ref{eqg}), by noting that in this case $p^2 = 0$, $p_v^2 = l_{\rm B}^2(V_0^2-2\Delta V_0)/ 2(\hbar \lambda v_{\rm F})^2$. Making use of $U(0,0;z) = U(0,1;z) =1$,  we have 

\begin{figure}[th!]
   \centering      
   \includegraphics[width=1.12\columnwidth]{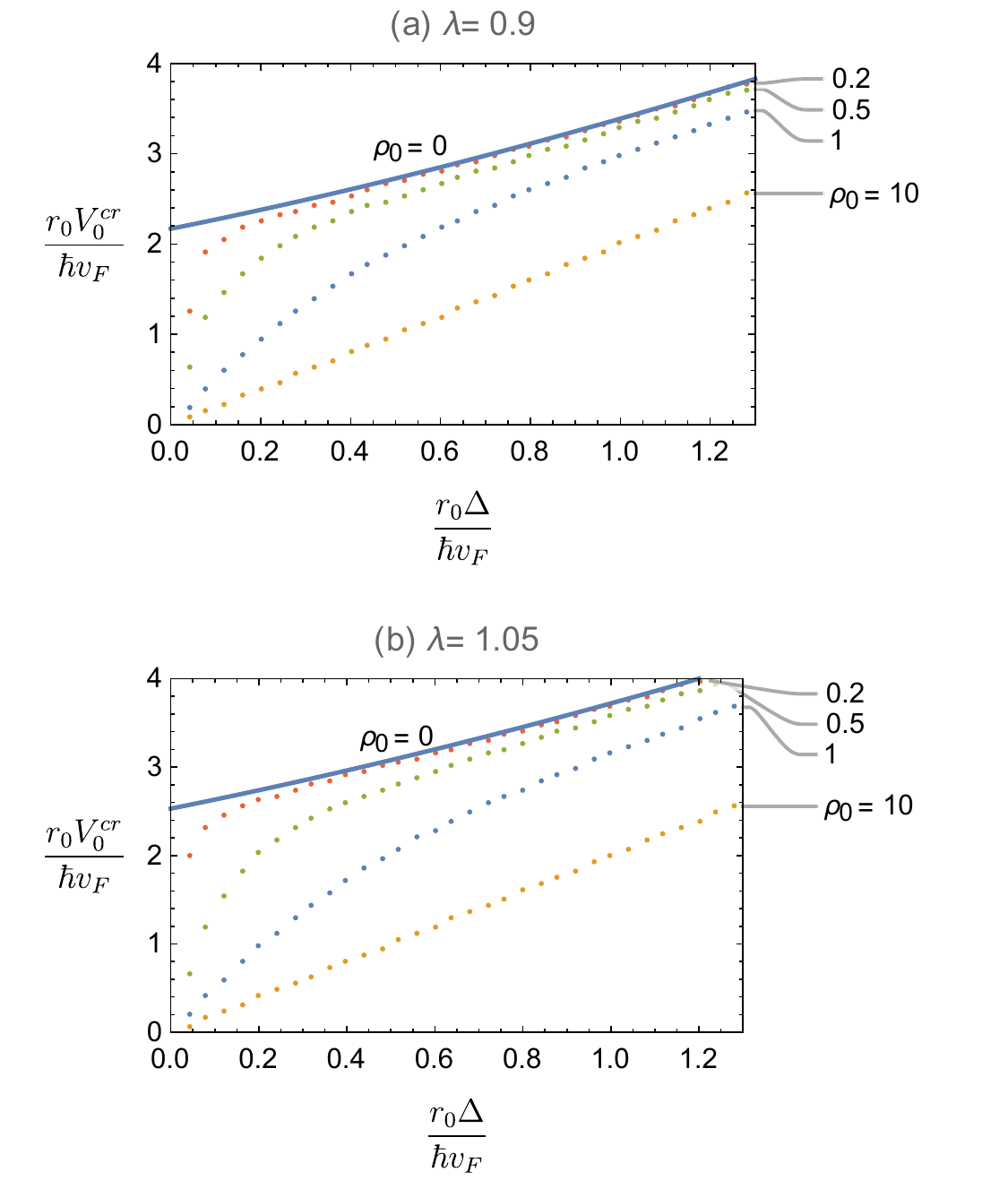}
   \caption{$V_{0}^{\rm cr}$ as function of $\Delta$ for different values of $\rho_0$ ($B_0$) and including the strain parameter $\lambda$ with the extreme values (a) $\lambda = 0.9$ and (b) $\lambda = 1.05$, representing maximum compression and expansion of the LLs respectively. Here,  $\rho_0 =0$ corresponds to  $B_0=0$.}
    \label{fig:v_0_r_l}
\end{figure}

\begin{equation}
   V_{0}^{\rm cr} = 2\Delta \left[ 1+ \frac{2\ M\left(a,1;\rho_0^2/2\right)}{\rho_0^2\ M\left(1+a,2;\rho_0^2/2\right)} \right] ,
\end{equation}\\
where $a=l_{\rm B}^2\ V_{0}^{\rm cr}(V_{0}^{\rm cr}-2\Delta)/ 2(\hbar \lambda v_{\rm F})^2$. 
The critical value potential  as a function of the quasiparticle mass gap is depicted in Fig.~\ref{fig:v_0_r_l} for different values of the parameters $\rho$ and $\lambda$. It is clearly seen that  $V_{0}^{\rm cr}$ decreases drastically as the magnetic field strength $B_0$ increases (because $\rho_0 = r_0/l_{\rm B}$ and $l_{\rm B} \sim \sqrt{1/|B_0|}$) for fixed values of $r_0$ and $\Delta$.


From the physical point of view, it represents the fact that the external magnetic field $B_0$ brings electronic orbitals closer to the nucleus, and thus, the attraction between an orbiting electron and a nucleus gets more intense, lowering the critical value of the potential. In addition, the value of the parameter $\lambda$, which contains information about the strain, provokes the critical potential value $V_{0}^{\rm cr}$ to grow for $\lambda>1$  or to decrease for $\lambda<1$ (Fig.~\ref{fig:v_0_r_l}), for fixed values of $r_0$, $\Delta$ and $B_0$. This result leads us to conclude the possibility to induce such an strain in the graphene lattice which makes electronic collapse easier to take place in one direction and more difficult in the other, being capable to control an effective directed-current. Another observation is that in all the cases with $B_0 \neq 0$, $V_{0,cr}$ goes to zero as $\Delta$ decreases arbitrarily. From that, we can conclude that the presence of an homogeneous magnetic field catalyzes atomic collapse in the Coulomb center doped graphene for any strain. In particular, in the case of gapless graphene in the presence of an homogeneous magnetic field atomic collapse would take place for any value of $Z$.

\section{Point dipole in strained graphene}

In this section we address the case of a point electric dipole placed  in a graphene plane~\cite{de2014electric}, but considering additionally the effect of uniform uniaxial strain.
The electric dipole consists in two opposite charges $+Q$ and $-Q$ separated a distance $d$. Thus, its potential taking strain into account is~\cite{jackson2007classical}
\begin{equation}
V_{\rm d}(r',\phi) = \frac{p/d}{r'_+}-\frac{p/d}{r'_-},
\end{equation}
where $p=Qd$ is the dipole moment, and  
\begin{equation}
    r'_{\pm} = \sqrt{\left(x\pm d/2\right)^2 + y^2}
\end{equation}
\noindent are the lengths of the vectors $\mathbf{r'}_+$, $\mathbf{r'}_-$ under strain joining the negative and positive charge positions with the observation point respectively.  From Eq.~(\ref{transform})
\begin{align}
\dfrac{\zeta^{1/2}r}{r'_{\pm}} &= \left[ 1 + c_1 \sin^2 \phi \pm \zeta^{-1/2} \dfrac{d}{r} \cos \phi + \left(\dfrac{c_2}{r}\right)^2 \right]^{-1/2} \nonumber \\ \nonumber \\ &\approx 1- \dfrac{1}{2} \left[ c_1 \sin^2 \phi \pm \zeta^{-1/2} \dfrac{d}{r} \cos \phi + \left(\dfrac{c_2}{r}\right)^2 \right], 
\end{align}
with $c_1=\zeta^{-2}-1$ and $c_2= d/2$. So that
\begin{align}
\dfrac{r}{r'_+} - \dfrac{r}{r'_-} &= -\zeta^{-1} \dfrac{d}{r}\cos \phi,
\end{align}
and thus, in the limit $d\to 0$, the point electric dipole potential under strain is
\begin{equation}
V_{\rm d}(r', \phi)\approx -\zeta^{-1} \dfrac{p\cos \phi}{r^2}= - \frac{p'\rm{cos} \phi}{r^2} = V_{\rm d}(r, \phi),
\end{equation}
with the redefinition of the dipole moment $p'=\zeta^{-1}p$.
%
%
The corresponding Hamiltonian has the form
\begin{equation}
    H= \hbar\ \tensor{v}_{\rm F} \cdot \boldsymbol{\sigma} \cdot \mathbf{p} + \Delta \sigma_z + V_{\rm d},
\end{equation}
where we are introducing the strain effects by the tensor Fermi velocity as in the previous section.
This Hamiltonian  has an intrinsic electron-hole symmetry $UHU^{\dagger}= -H$, with the unitary operator $U=\sigma_x R_x$ satisfying $U^2=1$, where $R_x$ is the operator of reflection $x\rightarrow -x$. An eigenstate $\Psi_E(x,y)$ with energy $E$ is mapped to another eigenstate with energy $-E$ as
\begin{equation}
    \Psi_{-E}(x,y) = U\Psi_E(x,y) = \sigma_x \Psi_E(-x,y).
\end{equation}
Hence, all solutions of the Dirac equation come in pairs with $\pm E$, and it is enough to study one of the eigenstates to automatically know the other.

Thus, the Dirac equation governing the system is written as
\begin{equation}
   \left( \begin{matrix} V_{\rm d}(r,\phi) + \Delta - E & \hbar v_{\rm F}(aP_x -ibP_y) \\  \\ \hbar v_{\rm F}(aP_x +ibP_y)  & V_{\rm d}(r,\phi) - \Delta - E \end{matrix} \right) \Psi =0,
\end{equation}
where $P_x = -i\partial_x$ and $P_y = -i\partial_y$ are the $x,y$ components of the linear momentum, and $\Psi = \left( \psi_A \quad \psi_B  \right)^T$ . 
Thus, the Dirac equation turns into
\begin{equation}
   \left( \begin{matrix} V_{\rm d}(r,\phi) + \Delta - E & \hbar \lambda v_{\rm F} {\rm e}^{-i\phi}  \left( -i \partial_r - \frac{\partial_{\phi}}{r} \right) \\ \\ \hbar \lambda v_{\rm F} {\rm e}^{i\phi}  \left( -i \partial_r + \frac{\partial_{\phi}}{r} \right) & V_{\rm d}(r,\phi) - \Delta - E \end{matrix} \right) \Psi =0.\label{diracdipole}
\end{equation}

Considering energies near the threshold $-\Delta$, $E=-\Delta + \epsilon$, with $|\epsilon|\ll \Delta$, Eq.~(\ref{diracdipole}) is written as
\begin{equation}
    \left( \begin{matrix} \frac{-p' \rm{cos}\phi}{r^2}+2\Delta -\epsilon & \hbar\lambda v_{\rm F} {\rm e}^{-i\phi}  \left( -i \partial_r - \frac{\partial_{\phi}}{r} \right) \\ \\ \hbar \lambda v_{\rm F} {\rm e}^{i\phi}  \left( -i \partial_r + \frac{\partial_{\phi}}{r} \right) & \frac{-p' \rm{cos}\phi}{r^2} -\epsilon \end{matrix} \right) \Psi =0. 
\end{equation}
For $p'\ll d^2\Delta$,  we have
\begin{equation}
    \psi_A \approx \frac{\hbar\lambda v_{\rm F}}{2\Delta} {\rm e}^{-i\phi}  \left( i \partial_r + \frac{\partial_{\phi}}{r} \right) \psi_B,
\end{equation}
and thus we are lead to the equation for $\psi_B$ of the form
\begin{equation}
\left(-\frac{\hbar^2\lambda^2 v_{\rm F}^2}{2\Delta} \nabla^2 + \frac{p' \rm{cos}\phi}{r^2} +\epsilon\right) \psi_B =0,
\end{equation}
where $\nabla^2 = \partial_r^2 + \frac{1}{r} \partial_r + \frac{1}{r^2} \partial_{\phi}^2$ is the 2D Laplace operator in cylindrical coordinates. 
Let us propose a separable solution of the form
\begin{equation}
    \psi_B(r,\phi) = R(r) \Phi(\phi),
\end{equation}
which upon separation of variables can be written as
\begin{align}
    & \qquad  \frac{r^2}{R(r)}\frac{\partial^2 R(r)}{\partial r^2} - \frac{2\Delta}{\hbar^2\lambda^2 v_{\rm F}^2} r^2 \epsilon = \nonumber \\ & - \frac{1}{\Phi(\phi)} \frac{\partial^2 \Phi(\phi)}{\partial \phi^2} + \frac{2\Delta}{\hbar^2\lambda^2 v_{\rm F}^2} p' \rm{cos}\phi = \gamma.
\end{align}
where $\gamma$ is the separation constant.
The angular equation
\begin{equation}
    \left[ \frac{d^2}{d\phi^2} - \frac{2\Delta}{\hbar^2\lambda^2 v_{\rm F}^2} p' \rm{cos}\phi + \gamma\right] \Phi(\phi) =0,
\end{equation}
which is independent of the parameter $\epsilon$, corresponds to a Mathieu equation~\cite{gradshteyn2014table, abramowitz1948handbook}, whose solutions
\begin{equation}
    Y_{j,\kappa}(\phi) = \begin{cases} {\rm ce}_{2j}\left(\frac{\phi}{2}, \frac{4p'\Delta}{\hbar^2\lambda^2 v_{\rm F}^2}\right), \text{ for } \kappa= +, \\ \\
    {\rm se}_{2j}\left(\frac{\phi}{2}, \frac{4p'\Delta}{\hbar^2\lambda^2 v_{\rm F}^2}\right), \text{ for } \kappa= -, \end{cases}
\end{equation} 
where $\kappa= \pm$ is the parity, are $2\pi$-periodic and are only permitted for characteristic values of $\gamma$,
\begin{equation}
\gamma_{j,\kappa}(p) = \begin{cases} \frac{1}{4} a_{2j}\left( \frac{4p'\Delta}{\hbar^2\lambda^2 v_{\rm F}^2}\right), \text{ for } \kappa= +, \\ \\  \frac{1}{4} b_{2j}\left(\frac{4p'\Delta}{\hbar^2\lambda^2 v_{\rm F}^2}\right), \text{ for } \kappa= -, \end{cases}
\end{equation} 
where the angular momentum $j=0, 1, 2,\ldots$ is unconventional because of the anisotropy, satisfying the relation $j+\kappa \geq 0$. These characteristic values are ordered for a given value of $p'$ as $\gamma_{0,+}(p') < \gamma_{1,-}(p') < \gamma_{1,+}(p')<\gamma_{2,-}(p')<...$

The radial equation is given by
\begin{equation}
    \left[ \frac{d^2}{dr^2} - \frac{2\delta}{\hbar^2\lambda^2 v_{\rm F}^2} \epsilon + \frac{1}{r} \frac{d}{dr}- \frac{\gamma}{r^2}\right] R(r) =0,
\end{equation}
and has the form of a MacDonald equation~\cite{gradshteyn2014table}
\begin{equation}
    u'' + \frac{1}{z} u' - \left( 1-\frac{\nu^2}{z^2} \right) u = 0,
\end{equation}
whose solutions are the modified Bessel functions $K_{\nu}(z)$, which in our case correspond to $\nu=\sqrt{\gamma}$ and $z= (r\sqrt{2 \Delta \epsilon})/(\hbar\lambda v_{\rm F})$. Recall that for bound states ($\epsilon>0$), these solutions have to decay for $r \rightarrow 0$ and to be regular at the origin. For this purpose, we use the regularization condition $R(r_0)=0$, which gives the energy quantization condition within each tower of $(j,\kappa)$,
\begin{equation}
    \epsilon_{n,j,\kappa} = \frac{z_n^2\hbar^2 \lambda^2 v_{\rm F}^2}{2\Delta r_0^2},
\end{equation}
where $z_n$ are the zeros of the function $K_{\sqrt{\gamma_{j,\kappa}}}(z)$ satisfying $z_1<z_2<z_3<...$ Because this function only has zeros for $\sqrt{\gamma_{j,\kappa}}$ imaginary~\cite{gradshteyn2014table}, then the condition $\gamma_{j,\kappa}(p') <0$ is required for bound states. This is satisfied for values $p'> p_{j,\kappa}$ with
\begin{equation}
    \gamma_{j,\kappa}(p_{j,\kappa})=0.
\end{equation}

\begin{figure}[th!]
   \centering      
   \includegraphics[width=1.12\columnwidth]{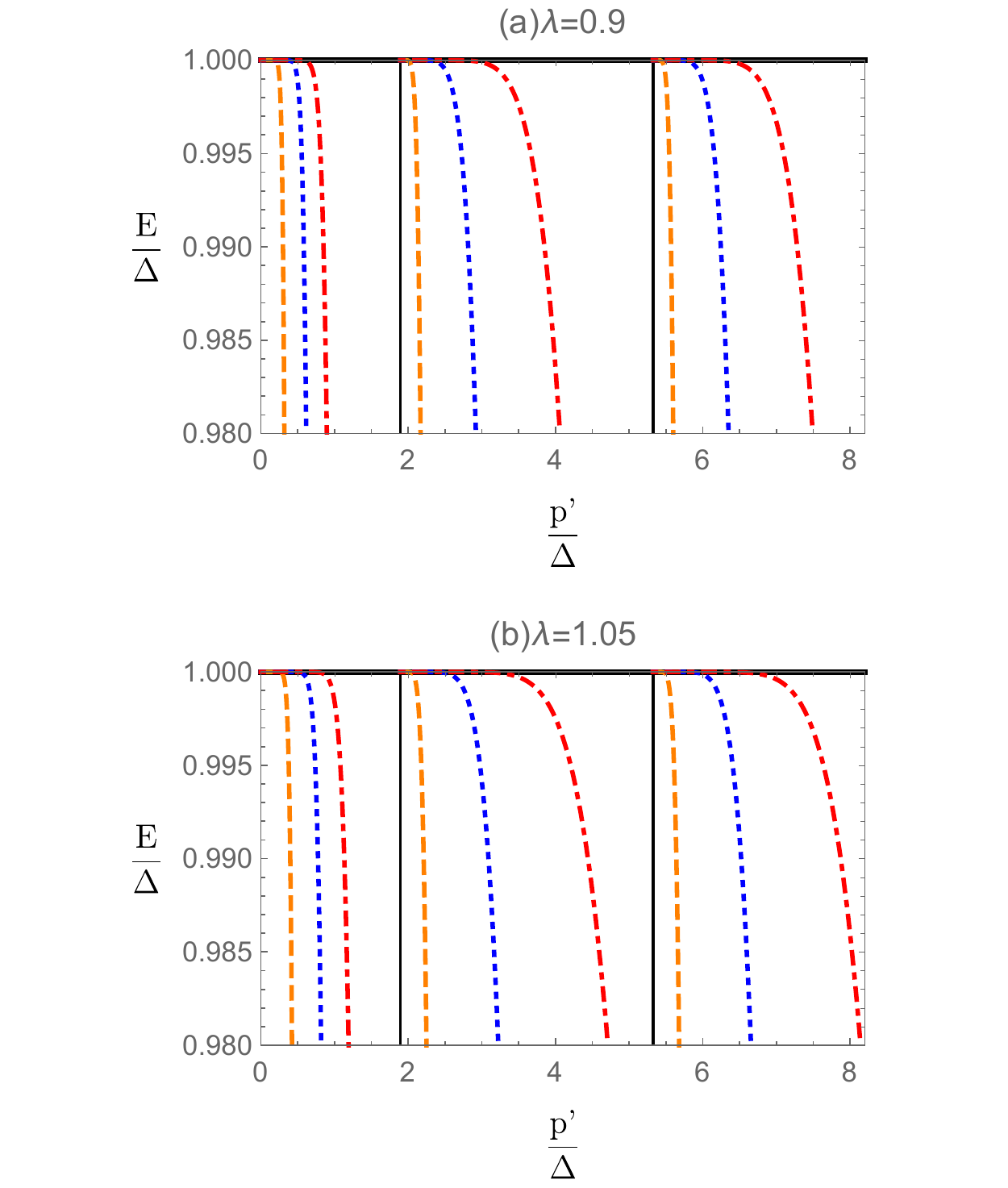}
   \caption{Energy near the edge $+\Delta$ of the first electron bound states with respect to the strained electric dipole moment for different values of the strain parameter (a) $\lambda=0.9$ and (b) {$\lambda=1.05$}.}
    \label{fig:dipolo_l}
\end{figure}

\noindent These values of $p_{j,\kappa}$ have been already presented in~\cite{de2014electric}.
%
%
By increasing the value of $p'$, each time that $p'$ hits a critical value $p_{j,\kappa}$, a new infinite tower of bound states emerges from the continuum. Then, by expanding $K_{is}(z)$ for small $z$ as in Eq.~(\ref{Besselz0}) (with $s_{j,\kappa}(p')= \sqrt{-\gamma_{j,\kappa}(p')}$ for $p'>p_{j,\kappa}$), we obtain an explicit expression for the bound energies near the edges $\pm \Delta$ given by
\begin{equation}
    \epsilon_{n,j,\kappa} = \frac{2\hbar^2 \lambda^2 v_{\rm F}^2}{\Delta r_0^2} {\rm e}^{\phi(s_{j,\kappa})} {\rm e}^{-2\pi n /s_{j,\kappa}},
\end{equation}
where $\phi(s) = (2/s) \ {\rm arg}\,\Gamma(1+is)$, and the values $s_{j,\kappa}$ are~\cite{de2014electric}
\begin{equation}
    s_{j,\kappa}(p') = \begin{cases} \frac{\sqrt{2}p'\Delta}{\hbar^2\lambda^2 v_{\rm F}^2}, \qquad (j,\kappa) = (0,+), \\ \\
   \frac{\alpha}{\hbar\lambda v_{\rm F}} \sqrt{(p'-p_{j,\kappa})\Delta}, \qquad j>0, \end{cases} 
\end{equation}
where $\alpha \approx 0.956$. It can be seen that these states verify  an Efimov-like universal scaling law as in~\cite{de2014electric}
\begin{equation}
    \frac{\epsilon_{n+1}}{\epsilon_n} = {\rm e}^{-2\pi /s_{j,\kappa}},
\end{equation}
but with the $\lambda$-dependence inserted in the $s_{j,\kappa}$ terms. Plots of the energy spectrum versus strained dipole moment $p'$ for different values of the strain strength $\lambda$ are presented in Fig.~\ref{fig:dipolo_l}. In all of them the first electron, $\epsilon_{n,j,\kappa}$, bound states are presented for each tower, decaying from the continuum to discrete values. 


Just as in the case of a single charged impurity seen in the previous section, the role of the strain strength is to promote (for values of $\lambda <1$) or inhibit (for values of $\lambda >1$) the decay of the electron and hole bound states, making the collapse in both systems easier or more difficult to occur depending on the strain. Also, for $n\rightarrow \infty$ we have that $\epsilon \rightarrow 0$, so that the edges $\pm \Delta$ are accumulation points. These facts suggest that electrons can be captured by a dipole potential in graphene, and strain can be used to tune the strength of confinement and to promote or inhibit atomic collapse.

\section{Conclusions}

In this article we have carried out an analysis of the bound states for the anisotropic 2D Dirac equation (by virtue of uniform uniaxial strain) in the potential of point electric monopole and dipole impurities in the context of graphene physics. In the former case, in absence of strain, it is known that the combined effect of a uniform magnetic field and a Coulomb impurity yield to a scenario of atomic collapse. Analytically, this problem cannot been solved. Nevertheless, by regularizing the Coulomb potential at the origin through a potential well, the problem exhibits interesting features, like atomic collapse in absence of magnetic fields provided the potential well has a value above certain critical number strongly connected with the size of the impurity. The magnetic field drives this critical strength to zero, such that atomic collapse occurs for impurities of arbitrary size. Furthermore, as the magnetic field increases its strength, more and more states with angular momentum $j<-1/2$ start to dive into the continuum. On the other hand, for the point dipole impurity, even though no collapse of states is observed, the appearance of cascades of infinitely many bound states with a universal Efimov-like scaling appears as a reminiscent of the true collapse that happens in the case of a finite-size dipole. Because the role of strain is seen in a renormalization of the Fermi velocity $v_{\rm F}\to \lambda v_{\rm F}$, for $\lambda<1$ these two scenarios are promoted to occur under less restrictive conditions, as compared with the ideal case. For $\lambda>1$, the situation is the opposite. As an additional result, we observe that the effect of strain in the Coulomb and point electric dipole potentials addressed in this work can be understood as a redefinition of the electric charge and the dipole moment, respectively, in terms of the strain parameter $\zeta$. 

The role of position-dependent strain is currently under consideration in our group, with the addition of the generation of electric fields due to this deformations. Along these lines, we point out the recent findings regarding the collapse of Landau levels under strain and a uniform electric field recently discussed in~\cite{Ghosh_2019}. Results will be presented elsewhere.

\section*{Acknowledgments}
The authors acknowledge the anonymous referees for their valuable comments to improve this work.

\section*{Data availability statement}
Data available on request from the authors.

\section*{Appendix}

Some useful formulae are presented in this Appendix. The Heaviside step function $\theta(x_0-x)$ found in the potential well owns the property
\begin{align}
    \theta[c(x_0-x)] &= \theta[x_0-x] \theta[c] + \theta[x-x_0]\theta[-c] \nonumber \\ &= \begin{cases} \theta(x_0-x), \ \ c>0, \\ \theta(x-x_0), \ \ c<0, \end{cases} \label{eq:heaviside}
\end{align}
with $c$ a scalar. In our case $c>0$, obtaining Eq.~(\ref{eq:well}).

Regarding properties of the $M(a,c;x)$ and $U(a,c;x)$ for $c<0$, we have that~\cite{bateman1953higher, gradshteyn2014table, abramowitz1948handbook}
\begin{align}
\lim_{c \to -m} \frac{M(a,c;x)}{\Gamma(c)} &= \frac{\Gamma(a+m+1)}{\Gamma(a)(m+1)!} x^{m+1}\nonumber\\
& \hspace{-1.5cm}\times M(a+m+1,m+2;x), \quad m=0, ~1, \ldots \\
U(a,c;x) &= x^{1-c} U(a-c+1,2-c;x),\label{eq:MU}
\end{align}
which along with the identity
\begin{equation}
\frac{\Gamma(a+1)}{\Gamma(a)} = a,\label{eq:MUG}
\end{equation}
allow to reach at the expressions in Eq.~(\ref{eqn}).

For the case of vanishing magnetic field, for the monopole impurity, we use the identities
\begin{align}
    &M(a,c;x) = {\rm e}^x U(c-a,c;-x),\\
   &\lim_{a\rightarrow \infty} M(a,c;x) = \Gamma (c)\ x^{\frac{1-c}{2}} J_{c-1}(2\sqrt{x}),\\
   &\lim_{a\rightarrow \infty}\left[\Gamma (1+a-c) \ U(a,c;-x/a)\right] =\nonumber\\
   &-i\ \pi {\rm e}^{i\pi c} x^{\frac{1-c}{2}} H^{(1)}_{c-1}(2\sqrt{x}),\ \ \mbox{Im}(x)>0,\label{eq:MUB0}
\end{align}
and assuming $|{\rm arg}\ a|<\pi$ for the last two expressions.

Some identities involving the Bessel functions are
\begin{align}
J_{-n}(z) &= (-1)^n J_n(z),\nonumber\\
 H^{(1)}_{-n}(z) &= (-1)^n H^{(1)}_n(z), \nonumber\\
 J_{n}(iz) &={\rm e}^{i\pi n /2} I_n(z),\nonumber\\
    H^{(1)}_{n}(iz) &=\frac{2}{i\pi}{\rm e}^{-i\pi n /2} K_n(z),
    \label{Bessel}
\end{align} 
and relevant the asymptotic forms as the argument $z\to 0$ are
\begin{align}
\lim_{z\rightarrow 0} I_n(z) & \simeq \frac{1}{\Gamma(n+1)} \left( \frac{z}{2}\right)^n, \quad n\geq 0, \nonumber\\
\lim_{z\rightarrow 0} K_n(z) &\simeq 
\begin{cases}
    -{\rm ln}\left( \frac{z}{2}\right) - \gamma_E, & n=0, \\ 
    \frac{\Gamma(n)}{2} \left( \frac{2}{z}\right)^2, & n>0.
    \end{cases}\label{Besselz0}
\end{align}

\bibliography{bib}

\end{document}